\documentclass[12pt]{iopart}
\usepackage{graphicx}
\usepackage{bm}
\usepackage{iopams} 

\begin{document}

\title[(NH$_4$)$_2$S$_x$ passivation of the (311)A GaAs surface]{The effect of (NH$_4$)$_2$S$_x$ passivation on the (311)A GaAs surface and its use in AlGaAs/GaAs heterostructure devices}

\author{D~J Carrad$^1$, A~M Burke$^1$, P~J Reece$^1$, R~W Lyttleton$^1$, D~E~J Waddington$^1$, A Rai$^2$, D Reuter$^2$, A~D Wieck$^2$, A~P Micolich$^1$}
\address{$^1$School of Physics, University of New South Wales, Sydney NSW 2052, Australia}
\address{$^2$Angewandte Festk\"{o}rperphysik, Ruhr-Universit\"{a}t Bochum, Bochum D-44780, Germany}
\ead{adam.micolich@nanoelectronics.physics.unsw.edu.au}

\submitto{\JPCM}
\date{\today}

\begin{abstract}
We have studied the efficacy of (NH$_4$)$_2$S$_x$ surface
passivation on the (311)A GaAs surface. We report XPS studies of
simultaneously-grown (311)A and (100) heterostructures showing that
the (NH$_4$)$_2$S$_x$ solution removes surface oxide and sulfidizes
both surfaces. Passivation is often characterized using
photoluminescence measurements, we show that while (NH$_4$)$_2$S$_x$
treatment gives a $40-60\times$ increase in photoluminescence
intensity for the (100) surface, an increase of only $2-3\times$ is
obtained for the (311)A surface. A corresponding lack of
reproducible improvement in the gate hysteresis of (311)A
heterostructure transistor devices made with the passivation
treatment performed immediately prior to gate deposition is also
found. We discuss possible reasons why sulfur passivation is
ineffective for (311)A GaAs, and propose alternative strategies for
passivation of this surface.
\end{abstract}
\maketitle

\section{Introduction}
Surface effects increasingly influence transport in electronic
devices as they are reduced in size. Understanding the surface's
influence and devising methods for minimizing its impact on
electronic performance is a vital aspect of device
development~\cite{LebedevPSS02}. Semiconductor surfaces are often
non-ideal, featuring complex surface reconstructions and abundant
dangling bonds. The latter produce localized energy levels in the
surface band structure that can act as metastable trapping sites.
GaAs surface states pin the surface Fermi energy near the middle of
the band-gap causing surface recombination and difficulties in
making ohmic contacts. These present difficulties for GaAs-based
devices such as photovoltaic cells and bipolar
transistors~\cite{LiuAPL88}.

Chalcogenide-based passivation has long been investigated towards
reducing surface-state problems in III-V
semiconductors~\cite{LebedevPSS02}. A favored route for GaAs surface
passivation involves inorganic sulfides such as (NH$_{4}$)$_{2}$S
and Na$_{2}$S~\cite{SandroffAPL87, YablonovitchAPL87}, which remove
the native surface oxide and adsorb S onto the Ga and/or As surface
atoms. The objective of passivation is to covalently satisfy all Ga
and As dangling bonds so that the resulting surface states have
energies in either the conduction or valence bands, where they no
longer act as charge traps~\cite{LebedevPSS02, YablonovitchAPL87,
CohenAM00}. This ideal is difficult to achieve, and resulted in a
plethora of passivation chemistries and treatments in both the
liquid and gas phases~\cite{LebedevPSS02, BessolovSST98}. Preceding
work focussed on basic surfaces such as (100), (110) and (111) due
to their applications in device technologies; sulfide passivation of
more complex surfaces such as (311)A has not been previously
reported. We are interested in (311)A heterostructures as they are
an underpinning materials platform for experimental studies of
low-dimensional hole systems. Compared to electrons, holes in GaAs
have enhanced carrier-carrier interactions due to an increased
effective mass and a curious spin-$\frac{3}{2}$ nature due to strong
spin-orbit effects. This has resulted in much interest in hole
systems for studies of the metal-insulator
transition~\cite{HaneinPRL98, PapadakisSci99}, bilayer quantum Hall
effect~\cite{TutucPRL04, ClarkePRB05}, Land\'{e} $g$-factor
anisotropy~\cite{WinklerPRL00, DanneauPRL06, ChenNJP10}, anomalous
spin-polarization effects~\cite{WinklerPRB05}, $0.7$ plateau in
quantum point contacts~\cite{DanneauPRL08}, Berry's phase in
Aharonov-Bohm rings~\cite{YauPRL02} and the quantum dot Kondo
effect~\cite{KlochanPRL11}. Hole quantum dots are also of interest
for quantum computing applications due to a lower spin-decoherence
time~\cite{HeissPRB07}.

We present a study of the efficacy of sulfur passivation treatment
of the (311)A GaAs surface, motivated by our previous study on the
origin of gate hysteresis in field-effect transistor devices made on
p-type Si-doped AlGaAs/GaAs heterostructures~\cite{BurkePRB12}. Here
we use photoluminescence (PL) and x-ray photoelectron spectroscopy
(XPS) measurements to analyze the relative efficacy of sulfur
treatment on the (311)A and (100) GaAs surfaces. This is combined
with electrical measurements of Schottky-gated transistors made
using p-type AlGaAs/GaAs heterostructures and different sulfur
passivation treatments to determine the corresponding effect on gate
stability. We find that sulfur treatment of the (311)A surface
removes the native oxide and replaces it with a sulfide layer, as it
does for (100), but does not produce a consistent, corresponding
improvement in photoluminescence intensity. We suggest this arises
from the monovalent nature of the Ga dangling bonds at the (311)A
surface. Additionally, (NH$_4$)$_2$S$_x$ treatment causes a lack of
reproducibility in the gate characteristics of AlGaAs/GaAs
transistors, likely related to the instability of the As-S bond. We
offer potential strategies for improved passivation of the (311)A
surface and discuss some practical issues faced in translating
sulfur treatments, normally used on bare semiconductor surfaces, to
the polymer resist based fabrication typical for transistor devices.

\section{Methods}

The most common approach to III-V semiconductor surface passivation
is the use of aqueous (NH$_4$)$_2$S and (NH$_4$)$_2$S$_{x}$
solutions (see appendix~A). These solutions have been successfully
applied to device structures ranging from GaAs
metal-insulator-semiconductor field-effect transistors
(MISFETs)~\cite{JeongJJAP95} to InAs/GaSb photodiodes~\cite{LiAPL07}
to III-V nanowire devices~\cite{SuyatinNano07}. We considered
aqueous and alcoholic solutions of Na$_{2}$S, (NH$_{4}$)$_{2}$S and
(NH$_{4}$)$_{2}$S$_{x}$, but mostly restrict ourselves to aqueous
solutions due to resist compatability issues outlined in appendix~A,
and (NH$_{4}$)$_{2}$S$_{x}$, as the excess sulfur tends to make them
more efficacious~\cite{FanJJAP88, NannichiJJAP88}.

Two concentrations of aqueous (NH$_4$)$_2$S$_x$ solution were
investigated, denoted `weak' and `strong', made from common stock
solution prepared by adding $3$~mol/L of elemental sulfur (Aldrich)
to $20\%$ (NH$_4$)$_2$S in H$_2$O (Aldrich). The stock solution is
stirred for several hours until the sulfur is completely dissolved,
and stored in a light-free environment to prevent
photodecomposition. Strong treatment involves a $10$~min sample
immersion in $\sim 5$~mL stock solution. Weak treatment involves
$2$~min immersion in $\sim 5$~mL of a $0.5\%$ dilution of stock
solution in deionized water. In both cases, passivation solution was
heated to $40^{\circ}$C in a water bath~\cite{JeongJJAP95}. All
samples were etched in 31\% HCl:H$_2$O for $30$~s prior to
passivation. Prolonged exposure to light and air can result in
surface reoxidation~\cite{BesselovSemi98}; hence devices were stored
in the dark between passivation and any subsequent fabrication steps
or measurements. We focused on the weak treatment here because
etching of the GaAs is sometimes observed~\cite{NannichiJJAP88}. The
heterostructures used for the devices have only a thin GaAs cap
layer protecting the active AlGaAs and GaAs layers underneath and we
wanted to avoid the risk of exposing the AlGaAs, which would oxidize
rapidly in air, or generating significant surface roughness due to
etching by the (NH$_{4}$)$_{2}$S$_{x}$ treatment. We also tested the
strong treatment on our devices to check that insufficient treatment
solution concentration was not responsible for the lack of
passivation efficacy (see section~3.3).

Two types of GaAs substrates were used: epitaxially-grown Si-doped
AlGaAs/GaAs heterostructures were used for device fabrication and
XPS studies, while `bulk' GaAs substrates without epilayers were
used for the PL studies. We used bulk wafer for PL because epilayers
interfere with the surface PL signal. The bulk wafers are polished,
undoped (semi-insulating) GaAs supplied by AXT. These wafers were
also used as the substrate for nominally identical (100)/(311)A
AlGaAs/GaAs heterostructures (Bochum 13473/13516), custom grown
side-by-side in a single deposition using molecular beam epitaxy
(MBE). The active region consists of $650$~nm undoped GaAs, $35$~nm
undoped Al$_{0.34}$Ga$_{0.66}$As, $80$~nm Si-doped
Al$_{0.34}$Ga$_{0.66}$As and a $5$~nm undoped GaAs cap. These
matched heterostructures have majority carriers of opposite sign --
n-type for (100) and p-type for (311)A -- due to the amphoteric
nature of Si dopants in AlGaAs~\cite{WangAPL85}.

Pieces approximately $3 \times 4$~mm$^{2}$ were cleaved from the
host wafer and cleaned with acetone and 2-propanol. Samples prepared
for XPS/PL measurements underwent sulfur passivation as described
above, with samples typically measured within $30$~min of
passivation to avoid surface re-oxidation. For device studies, Hall
bars with a height of $130$~nm were defined by photolithography and
wet etching using a $2:1:20$ buffered HF:H$_{2}$O$_{2}$:H$_{2}$O
solution (the buffered HF was 7:1 NH$_{4}$F:HF). Ohmic contacts for
(311)A devices were formed by vacuum evaporation of a $150$~nm 99:1
AuBe film, followed by annealing at $490^\circ$C for $90$~s.
Schottky gates were defined photolithographically using AZ nLOF2020
photoresist. Sulfur passivation was performed between development
and vacuum deposition of $20$~nm Ti / $80$~nm Au gate metal.
Passivated samples were stored under deionized H$_{2}$O during
transfer to the evaporator, and exposed to air for $<5$ min before
the evaporator chamber reached vacuum ($< 1$~mTorr). Significant
surface reoxidation should not arise from such brief air
exposure~\cite{BesselovSemi98}, as confirmed by PL in appendix~B. We
expect the passivated surface to remain robust after metal
deposition as the gate metal protects the surface from light/air.
Table~1 lists the samples/devices studied along with the wafer type,
surface orientation and passivation treatment used.

\begin{table}
\centering \caption{Samples studied: Bulk GaAs wafer pieces used for
PL labeled B1 - B4, Heterostructures without devices used for XPS
labeled H1 - H4, Modulation-doped heterostructure devices used for
gate hysteresis studies labeled D1 - D6.}

\begin{tabular}{cccc}
\hline
Sample & Wafer & Surface & Passivation\\
\hline
B1 & bulk & (100)& strong\\
B2 & bulk & (100)& weak\\
B3 & bulk & (311)A & strong\\
B4 & bulk & (311)A & weak\\
H1 & 13516a & (100) & none\\
H2 & 13516a & (100) & weak \\
H3 & 13516b & (311)A & none\\
H4 & 13516b &(311)A & weak\\
D1 & 13516b & (311)A & none \\
D2 & 13473b & (311)A & weak \\
D3 & 13516b & (311)A & weak\\
D4 & 13473b & (311)A & weak  \\
D5 & 13516b & (311)A & strong \\
D6 & 13516b & (311)A & strong + anneal\\
\hline \hline
\end{tabular}
\end{table}

Room temperature PL and XPS measurements were used to compare the
action of the (NH$_4$)$_2$S$_x$ passivation solution on the (311)A
GaAs surface with what is known/expected for (100) GaAs surfaces.
The PL excitation was provided by a $488$~nm Ar ion laser (Coherent
Innova 70). Luminescence was coupled to a 0.27m grating spectrometer
(J/Y SPEX 270M) and recorded by CCD, giving a spectral resolution of
$3$~nm. The PL apparatus has $1$~mW incident power, giving
$0.49$~W/mm$^{2}$ intensity at focus. PL spectra are normalized to
the peak intensity of a corresponding unpassivated reference sample.
XPS measurements were performed using a ThermoScientific
ESCALAB250Xi system with a monochromated Al K$\alpha$ source ($h\nu
= 1486.68$~eV at $164$~W power). A $500$~nm spot size and
$90^{\circ}$ photoelectron take-off angle were used. The C(1s) peak
at $285.0$~eV is the binding energy reference for all measurements,
which are expected to be accurate to $\pm0.1$~eV. The Avantage
software package was used for XPS peak fitting.

Standard four-terminal ac lock-in techniques were used for the
electrical studies, which were performed at $T = 4.2$~K in liquid
He. The drain current $I_{d}$ was measured with applied source-drain
bias $V_{sd} = 100~\mu$V at a frequency $f = 73$~Hz. A dc gate bias
$V_{g}$ was applied using a Keithley $2400$ to enable
monitoring/limiting of the gate leakage current $I_{g}$.

\section{Results}

\subsection{XPS study of the effect of (NH$_{4}$)$_{2}$S$_{x}$ on the (311)A GaAs surface}

\begin{figure*}
\includegraphics[width=15cm]{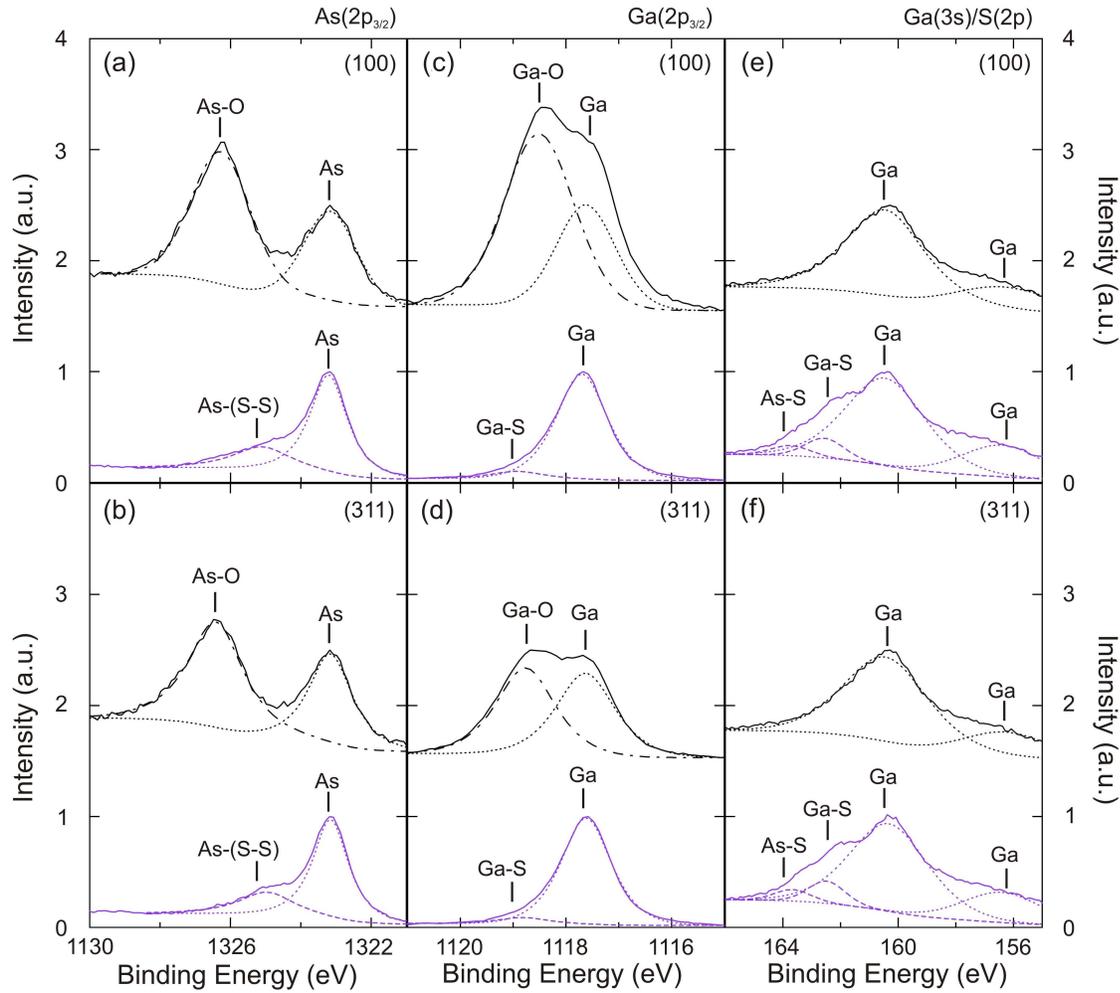}
\caption{(a,b) As(2p$_{3/2}$), (c,d) Ga(2p$_{3/2}$) and (e,f)
Ga(3s)/S(2p) XPS core level spectra for passivated (purple) and
unpassivated (black) GaAs surfaces with $(100)$ (top row) and
$(311)$A (bottom row) orientation. Intensity is normalized to the
Ga-As peak in each case, with unpassivated intensities offset
vertically by $1.5$ for clarity. Solid lines are the measured
spectra, dotted and dashed lines are peak fits for the Ga-As and X-Y
spectral peaks, where X = Ga,As and Y = S,O, respectively.}
\end{figure*}

We first discuss XPS studies of bare and passivated (311)A GaAs
surfaces. The surface sulfidization chemistry for bare and
passivated (100) GaAs surfaces is well characterized using
XPS~\cite{CarpenterAPL88, SandroffAPL89, SandroffJVSTB89,
SugaharaJAP91, PagetPRB96a, MoriartyPRB94, LuAPL93, ScimecaPRB92}.
We prepared and measured bare and passivated (100) reference
surfaces in parallel with (311)A surfaces to provide more direct
comparison, and to better isolate the effect of surface orientation
on passivation chemistry. XPS data for passivated and unpassivated
samples is shown in figure~1(a/c/e) and (b/d/f) for (100) and
(311)A, respectively. Beginning with the As(2p$_{3/2}$) core level
spectra in figure~1(a/b), the unpassivated surfaces show peaks
corresponding to GaAs ($1323.2$~eV) and As$_{2}$O$_{3}$
($1326.3$~eV) for both orientations. After passivation, the
As$_{2}$O$_{3}$ peak was eliminated and a small peak ($1324.5$~eV)
corresponding to disulfide bridges was
observed~\cite{SandroffJVSTB89}. The disulfide peak emerges to an
equivalent extent for (100) and (311)A. Turning to the
Ga(2p$_{3/2}$) spectra in figure~1(c/d), the unpassivated surfaces
show peaks corresponding to GaAs ($1117.4$~eV) and Ga$_{2}$O$_{3}$
($1118.5$~eV). The Ga$_{2}$O$_{3}$ peak intensity for (100) is $\sim
2\times$ that for (311)A. This likely reflects the single dangling
bond nature of surface Ga atoms on (311)A~\cite{ChadiJVSTB85,
StilesJVSTB85}. There is a clear Ga peak, and a weaker peak at
higher energy for the passivated surfaces that may correspond to
Ga-S bonding or residual Ga$_{2}$O$_{3}$. Unfortunately, the low
peak intensity makes conclusive assignment of these peaks' sources
difficult. Turning to the combined Ga(3s)/S(2p) spectra in
figure~1(e/f), the unpassivated surface gives two peaks at $156.4$
and $160.4$~eV corresponding to GaAs bonds. The passivated samples
show additional small peaks at $163.3$ and $162.1$~eV that are
likely As-S and Ga-S bonds (expected at $163.2$ and $162.3$~eV),
respectively~\cite{SakataJJAP94}. An alternate possibility is that
one of these peaks corresponds to a disulfide bridge (expected at
$163.5$~eV). Ultimately, the strong similarities between the XPS
spectra for (311)A and (100) suggest that the passivation treatment
removes surface oxide and establishes a surface sulfide layer with
roughly equivalent efficacy and chemistry.

\subsection{Comparative photoluminescence study of (NH$_{4}$)$_{2}$S$_{x}$ treated (100) and (311)A GaAs surfaces}

\begin{figure}
\includegraphics[width=8cm]{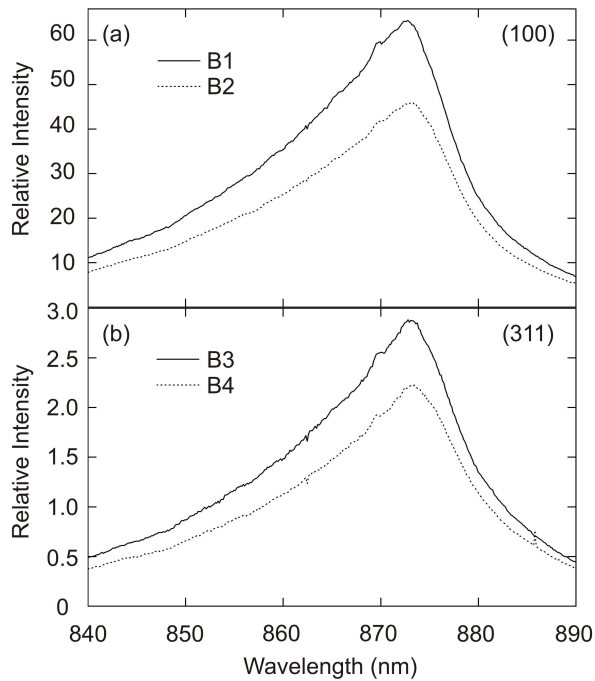}
\caption{PL intensity vs wavelength for (a) Samples B1 and B2 with
(100) surfaces passivated with strong (solid line) and weak (dotted
line) solution, respectively, and (b) Samples B3 and B4 with (311)A
surfaces passivated with strong (solid line) and weak (dotted line)
solution, respectively. The intensity is normalized to PL spectra
for an untreated sample with matching surface orientation.}
\end{figure}

A corresponding equivalence was not observed in the PL measurements.
Figure~2 shows PL spectra for (a) the (100) and (b) the (311)A
surfaces after passivation with the strong and weak
(NH$_{4}$)$_{2}$S$_{x}$ solutions. In each case, the intensity is
normalized to the peak intensity for an unpassivated surface with
matching orientation (see appendix~C), and represents the factor by
which passivation increases the PL intensity. Passivation increases
the PL intensity by only $2 - 3\times$ for (311)A GaAs, compared to
$40 - 65\times$ for (100) GaAs, suggesting passivation is much less
effective on (311)A GaAs. This likely arises from Ga, which presents
a double-dangling bond for (100) and a single-dangling bond for
(311)A, unlike As, which presents a double-dangling bond on both
surfaces~\cite{ChadiJVSTB85, StilesJVSTB85}. Although the Ga-S and
As-S XPS peaks show no difference between (100) and (311)A in
figures~1(e/f), the different Ga dangling bond valence for (311)A
may mean that sulfur atoms are unable fully satisfy all Ga dangling
bonds. This could explain the lower PL intensity observed for (311)A
in figure~2(b). We discuss this further in section~4.

The electrical data in section~3.3 is obtained at low temperature,
and the question could be asked: What about the PL at low
temperature? The difficulty is that PL does not just measure surface
recombination; it is also influenced by bulk radiative transitions
within the excitation photon penetration depth. Additional peaks
emerge in the PL at low $T$ corresponding to radiative transitions
not observed at $300$~K. These obscure the underlying band-to-band
PL peak used to assess surface recombination. Nevertheless, one can
compare the absolute PL intensity for passivated and unpassivated
surfaces. Skromme \etal~\cite{SkrommeAPL87} did this for (100) GaAs
at $T = 300$ and $1.8$~K. They observed $100 - 2800 \times$
increases in PL intensity at $300$~K upon passivation, dependent on
sample doping, but a $\sim3.5\times$ \emph{reduction} in passivated
sample PL intensity at $1.8$~K. Their ultimate conclusion was that
``recombination associated with the bare surface does not limit the
lifetime at low temperature as it does at $300$~K.'' In other words,
surface-states no longer dominate recombination at low $T$, other
recombination centers do. This does not mean that the surface-states
are removed, nor that they are no longer electrically active; it is
more that low $T$ PL is a poor probe of surface-states.

For completeness, we repeated the PL study for (311)A GaAs surfaces
at $T = 10$ and $20$~K: p-type AlGaAs/GaAs heterostructure devices
still exhibit qualitatively similar gate instability at these
temperatures~\cite{BurkePRB12}. The band-to-band recombination peak
intensity increased by $\sim4 \times$ for (311)A GaAs with the
weaker (NH$_4$)$_2$S$_x$ solution applied. As
Skromme~\etal~\cite{SkrommeAPL87} point out, such a small change is
generally considered negligible in a PL measurement. In particular,
the increase is small compared to the $40 \times$ increase in
band-to-band peak intensity obtained for passivated (100) GaAs
obtained at room temperature. Ultimately, our conclusion matches
Skromme {\it et al.}~\cite{SkrommeAPL87}: low $T$ PL is a poor probe
of surface-states. We instead turn to heterostructure devices to
further examine (311)A surface passivation.

\subsection{Effect of passivation on gate hysteresis in (311)A heterostructure devices}

\begin{figure*}
\includegraphics[width=14cm]{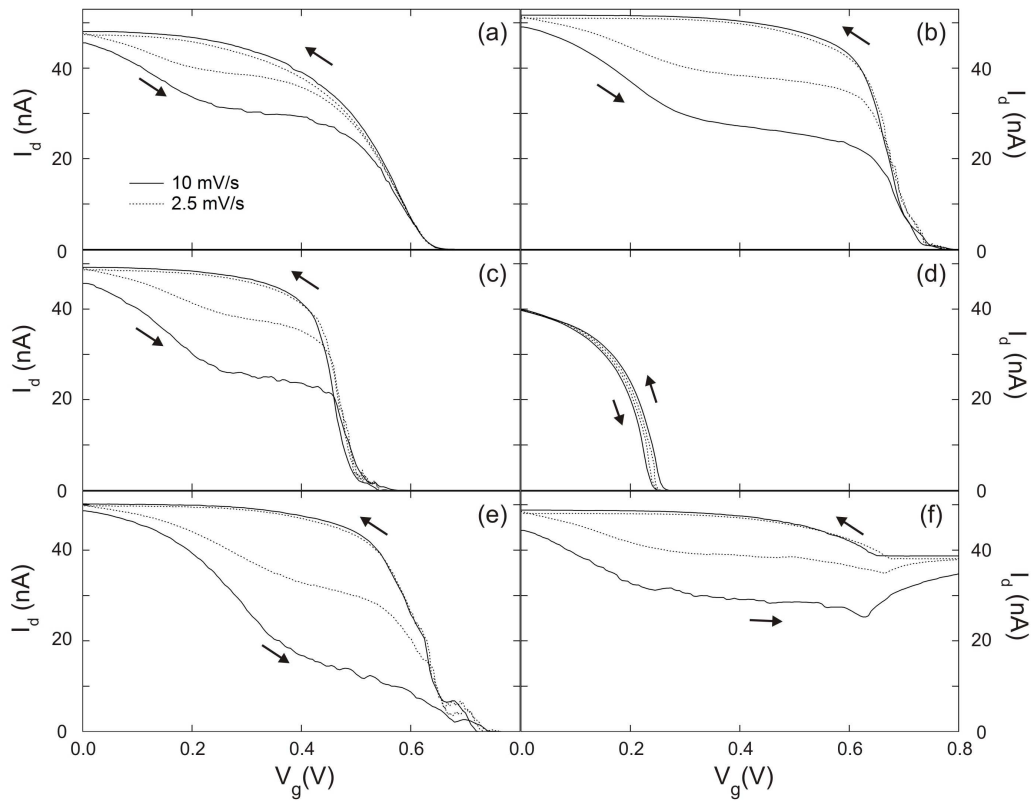}
\caption{Drain current $I_{d}$ vs gate bias $V_{g}$ for Devices (a)
D1 - unpassivated, (b-d) D2-D4 - weak passivation, (e) D5 - strong
passivation and (f) D6 - strong passivation and a post-passivation
anneal at $360^{\circ}$C for $10$~min under Ar. The arrows indicate
sweep direction (upsweep/downsweep); the solid (dashed) lines
indicate $V_{g}$ sweep rates of $10$($2.5$)~mV/s, respectively.
Pinch-off could not be achieved for Device D6; the gate leaks
strongly as $V_{g} \rightarrow +0.6$~V. The $I_{d}$ rise for
$V_{g}~>~+0.61$~V (upsweep), and $I_{d}$ plateau on return to
$V_{g}~=~+0.61$~V (downsweep) represent the current limiting action
of the $V_{g}$ source, which holds $V_{g}~=~+0.61$~V for all set
$V_{g}~>~+0.61$~V where $I_{g}~>~50$~nA.}
\end{figure*}

Gated Hall bars made using Si-doped (311)A AlGaAs/GaAs
heterostructures display strong gate hysteresis with an
anticlockwise hysteresis loop in the $I_{d}$ versus $V_{g}$
characteristics. This hysteresis is consistent with trapping of net
negative charge between the gate and the 2DHG, either in
surface-states or the modulation doping layer~\cite{BurkePRB12}.
Figure~3(a) shows a typical hysteresis loop obtained from an
unpassivated (311)A device. The most striking feature is the long
plateau at intermediate $V_{g}$ whilst sweeping to positive $V_{g}$.
Here, 2DHG depletion is strongly suppressed due to surface-state
charge trapping and/or dopant layer charge migration. Depletion
resumes for sufficiently positive $V_{g}$, and in most
Schottky-gated devices we studied, pinch-off (i.e., $I_{d} = 0$) is
attained before $V_{g}~>~+1$~V. The $I_{d}$ plateau length and
pinch-off voltage $V_{p}$ provide a measure of charge
trapping/migration within the device. Hence an effective surface
passivation should result in reduced $I_{d}$ plateau length and
lower $V_{p}$. Note that $I_{d}$ versus $V_{g}$ is a reasonable
approximation to a capacitance-voltage (C-V) study because $I_{d}$
depends upon the insulator capacitance. MOS capacitor C-V studies
are just a simpler route to studying insulator trapping that removes
the need for an FET conducting channel.

Figures~3(b-f) show electrical characteristics for passivated
devices D2-D6. The results for the weak passivation solution vary,
with the $I_{d}$ plateau extending (fig.~3(b)), shortening
(fig.~3(c)), and in one case, disappearing entirely (fig.~3(d)). The
behaviour in figure~3(b/c) is most typical; across the five devices
we observe $V_{p} = +0.25~-~+0.8$~V and plateau lengths up to
$0.4$~V. We have been unable to reproduce the outcome for device D4
(fig.~3(d)) in a second device despite numerous attempts; yet the
characteristics in fig.~3(d) were repeatable for subsequent
cooldowns of device D4. A more focussed surface chemistry study may
identify a route for producing the desirable outcome in fig.~3(d)
more consistently.

A possible argument for the inconsistent results in fig.~3(b-d) is
an insufficiently strong passivation treatment. Figure~3(e) shows
results from an attempt to more effectively passivate the GaAs
surface. Device D5 was prepared using the $200\times$ more
concentrated passivation solution, and although the $I_{d}$ plateau
at intermediate $V_{g}$ is weakened, there is little reduction in
$V_{p}$ and the device becomes more unstable at low $I_{d}$. This
suggests that the inefficacy of passivation treatment on the
hysteresis is not related to insufficient sulfidization.

Post-passivation annealing is an informative experimental tool
because the large difference in Ga-S and As-S bond
stability~\cite{HirayamaAPL89} means annealing increases Ga-S
bonding at the expense of As-S surface bonds~\cite{SugaharaJAP91,
SpindtAPL89a, SugaharaSurfSci91}. Little is known about the sulfur
chemistry of the (311)A surface, but since (311)A surface As atoms
are (100)-like (see section 4), we assume that desorption of As
surface bonds is complete for samples annealed above
$350^{\circ}$C~\cite{MoriartyPRB94, SugaharaSurfSci91, PagetPRB96b}.
Device D6 was prepared using the strong solution followed by a
post-passivation anneal for $10$~min at $360^{\circ}$C to determine
the influence As-S bonds have on the electrical characteristics. The
entire anneal process was conducted at $1$~atm Ar to prevent
oxidation, with photo-processing and metallization performed as soon
as practicable thereafter. The robustness of passivation to
subsequent photo-processing is discussed in appendix~B.

Post-passivation annealing appears detrimental for the (311)A
surface; as figure~3(f) shows, pinch-off cannot be achieved in these
devices. The $I_{d}$ plateau on sweeping to positive $V_{g}$ begins
at a lower $V_{g}$ and extends such that $I_{g}$ exceeds the $50$~nA
limit set by the gate voltage source before the end of the $I_{d}$
plateau (c.f. figure~2 of \cite{BurkePRB12}). The lack of depletion
in figure~3(f), particularly at low $V_{g}$, suggests that surface
As-X bonds may play an important role in the hysteresis. The
corollary is that As-S bonding may give reduced hysteresis, which
may explain the hysteresis-free behaviour in device D4. The
variability in passivation efficacy found in figures~3(b)-(d) may
also be symptomatic of low As-S bond stability~\cite{HirayamaAPL89,
SpindtAPL89a}, which leads to surface As accumulation and As-As and
As-O surface bond formation with H$_{2}$O washing~\cite{LuAPL93,
PagetPRB96b, BerkovitsJAP91}. This would make aqueous passivation
treatments a more capricious and variable prospect compared to
gaseous treatments, for example. Gates on (311)A that underwent
post-passivation annealing also tend to be more leakage prone. This
may arise from Be diffusion from the ohmic contacts; the contacts
are deposited and annealed prior to the passivation and
post-passivation anneal. Although the post-passivation anneal is
$130^{\circ}$C lower in temperature than the ohmic contact anneal,
its duration is more than six times longer. The rapid diffusion and
surface aggregation of Be in GaAs is well known~\cite{IvanovJCG91}.

Returning briefly to device D4, one possibility is that passivation
has produced a sharp reduction in the low energy tail of the
surface-state spectrum or the spectrum has changed such that a large
subset of the surface-states local to the surface Fermi energy of
the untreated surface was shifted in energy (i.e., passivation
induces a change in band-bending). In both cases, reduced charge
trapping at low $V_{g}$ would cause rapid depletion, allowing
pinch-off before onset of the $I_{d}$ plateau. A partially effective
passivation may be sufficient to achieve this. Device D4, although a
one-off example, indicates that a passivation treatment that can
significantly reduce the gate hysteresis in (311)A heterostructure
devices may exist.

\section{Discussion}
Drawing together the XPS, PL and electrical measurements above, it
is evident (NH$_4$)$_2$S$_x$ treatment removes oxide and sulfidizes
the surface for both orientations. The PL intensity enhancement is
smaller for (311)A, and there is little corresponding improvement in
gate hysteresis for (311)A-based FET devices. We embarked on this
study expecting that sulfur passivation may lessen the gate
hysteresis; we are instead left with several questions: Why does
(NH$_4$)$_2$S$_x$ produce a clear chemical change for (311)A with
little improvement in PL intensity and gate characteristics? Is
there something about the (311)A surface that would make this
expected behaviour? Is there some insight for how a more effective
passivation treatment might be formulated? We now attempt to answer
these questions.

Higher-Miller-index surfaces present as linear combinations of
lower-Miller-index surfaces. The (311) surface can be considered the
average of (100) and (111) surfaces.~\cite{WangAPL85, ChadiPRB84}
The (311)A surface studied here ideally presents equal densities of
(100)-like As double-dangling bonds and (111)A-like Ga
single-dangling bonds. We propose that the (311)A surface's
bimolecular nature is central to the inefficacy of sulfur
passivation despite the clear sulfur binding evident by XPS
(figure~1). For clarity of later discussion, we first briefly
address how (NH$_{4}$)$_{2}$S$_{x}$ treatment affects surface Ga and
As atoms for (100) and (111)A.

(NH$_{4}$)$_{2}$S$_{x}$ treatment produces a surface containing
Ga-S, As-S and surface-bound S-S dimers; however due to the reduced
stability of the As-S bond~\cite{HirayamaAPL89, SpindtAPL89a}, a
H$_{2}$O rinse leaves mostly Ga-S bonds with remaining surface-bound
S forms washed away~\cite{LuAPL93, BerkovitsJAP91}. Thermal
annealing exacerbates the dominance of Ga-S
bonds~\cite{SugaharaJAP91, PagetPRB96a, SpindtAPL89a}, making the
Ga-S bond the logical first consideration. For (100), the Ga-bound S
atom adopts a Ga-S-Ga bridge configuration~\cite{SugaharaJVSTA93,
SugiyamaPRB94} to satisfy the surface Ga double-dangling bond. Total
energy calculations using density-functional theory suggest the
resulting bonding and antibonding orbitals for Ga-S bonds sit within
the valence and conduction bands~\cite{OhnoPRB90}. This should
produce a substantial reduction in Ga-related mid-gap surface-state
density, and depinning of the surface Fermi level. Indeed, a
substantial reduction in Ga-related surface state density close to
the conduction band was reported for capacitance-voltage (C-V) and
deep level transient spectroscopy (DLTS) studies of passivated (100)
surfaces~\cite{LiuAPL88, JeongJJAP95, FanJJAP88}. In contrast, for
(111)A, the Ga-bound S atom sits above the surface Ga atom to
satisfy the surface Ga single-dangling bond~\cite{SugiyamaAPL92,
SugiyamaPRB93, MurphySurfSci94}. Calculations suggest the bonding
and anti-bonding orbitals sit inside the band-gap in this
case~\cite{OhnoPRB91}. Hence Ga-related mid-gap states for
unpassivated (111)A are replaced by Ga-S states nearer the valence
band for passivated (111)A. This should cause the surface Fermi
energy to pin closer to the valence band maximum rather than
depinning~\cite{OhnoPRB91}. That said, experiments suggest the
surface Fermi energy moves away from the valence band maximum upon
(111)A passivation~\cite{MurphySurfSci94}, possibly due to S
substituting some uppermost sub-surface As atoms in addition to
bonding to surface Ga. Either way, passivation efficacy for (111)A
would be diminished compared to (100)~\cite{MurphySurfSci94}.

The behaviour of surface As is more difficult because although
surface As for (311)A is (100)-like and passivation for (100) has
been heavily studied, the role of As-S bonding in passivation for
(100) remains poorly understood. For example, some theoretical
studies suggest the As-S antibonding state for (100) sits within the
band-gap~\cite{OhnoPRB90, OhnoPRB91}, while others place these
states within the valence band~\cite{RenPRB90}. The role and
importance of As-S bonding is also debated on the experimental side.
Some suggest As-S and As-(S-S) are central to
passivation~\cite{SandroffAPL89, MoriartyPRB94, FanJJAP88b}, others
that Ga-S bonds are key~\cite{LuAPL93, SugaharaSurfSci91,
PagetPRB96b}, and some that both Ga-S and As-S bonds are
involved~\cite{SikJECS97}. One difficulty is the relative As-S bond
weakness and tendency for surface As accumulation, which causes
additional mid-gap levels~\cite{PagetPRB96b}.

We now consider (311)A specifically: In essence, (311)A is a worst
case scenario. First, the Ga dangling bonds are (111)-like and Ga-S
bonding should produce states within the band-gap~\cite{ChadiPRB84,
OhnoPRB91}. This could explain the lack of PL intensity increase for
the sulfidized (311)A surface. Second, (311)A should display equal
proportions of Ga and As dangling bonds~\cite{WangAPL85,
ChadiPRB84}. This means that (311)A surface passivation will always
be difficult; compromised by the As-S bond's weakness to H$_{2}$O
exposure~\cite{LuAPL93, PagetPRB96b, BerkovitsJAP91}. An action as
simple as changing the rinse time could alter the As-S bonding,
giving different gate characteristics to each device, as in fig.~3.
Finally, the (311)A surface can display metastable
reconstructions~\cite{TaguchiPRB05}; this may further complicate the
surface chemistry and electronic states. This explanation for the
passivation inefficacy of (311)A is simplistic, further theoretical
and surface studies would significantly enhance understanding.
Studies of (311)A surface-state spectrum using frequency dependent
C-V~\cite{FanJJAP88, SikJECS97} or DLTS~\cite{LiuAPL88} would also
be valuable.

The final question is whether a formulation exists that might
passivate (311)A more effectively. The problem here is that both the
Ga and As surface atoms need attention. The problem with (311)A
surface Ga is that the dangling bond is monovalent and sulfur is
divalent. A monovalent adsorbate, e.g., Cl, is one alternative. HCl
passivation of (111)A has been demonstrated~\cite{LuAPL95}, but we
found no improvement for (311)A heterostructure devices by HCl
treatment. Recent studies have shown that post-chloridation thermal
annealing, possibly combined with hydrazine treatment, can
significantly enhance Cl passivation efficacy by removing excess
As~\cite{TraubJACS08}. Further studies of more complex Cl treatments
for (311)A would be of interest. Turning now to dangling As bonds,
an obvious alternative is Se, with several reports that it is more
effective and stable against oxidation than S
treatment~\cite{SandroffJAP90, KuruvillaJAP93}. The greater
subsurface penetration of Se~\cite{ScimecaPRB92, ScimecaAPL93,
PashleyJVSTA94} may also be favorable assuming an As antisite defect
model~\cite{SpindtAPL89b} for GaAs surface states. Possible
formulations include Na$_{2}$Se in NH$_{4}$OH followed by
Na$_{2}$S$_{(aq)}$~\cite{SandroffJAP90}, SeS$_{2}$ in
CS$_{2}$~\cite{KuruvillaJAP93} or Se-loaded
(NH$_{4}$)$_{2}$S~\cite{BelkouchSSE96}. Note that this would only
deal with surface As; passivation of (111)-like surface Ga bonds
would entail additional treatment. Ultimately, As-chalcogen bond
stability and surface As accumulation may still be an issue even
with Se-passivation.

Engineering of the heterostructure's cap layer may be the most
viable alternative. This would involve growing a degenerately-doped
GaAs cap layer~\cite{BurkePRB12}. This cap could be partitioned by
wet-etching to form gates for the device, as in undoped
Heterostructure Insulated Gate Field Effect Transistors
(HIGFETs)~\cite{KaneAPL98, ClarkeJAP06}, although the device would
still operate as `normally on' due to the Si modulation doping
layer. The advantage is that the the highly-conductive cap screens
the conducting channel from the surface-states; the disadvantage is
that it requires far more complex device processing.

We finish by commenting on the implications for C-doped
(100)-oriented AlGaAs/GaAs heterostructures~\cite{GrbicAPL05,
GrbicPRL07, KomijaniEPL08, CsontosAPL10, KomijaniEPL10}, where gate
hysteresis is also observed. We expect the hysteresis in these
heterostructures to arise mainly from dopant
fluctuations~\cite{BurkePRB12}. We observe no hysteresis in (100)
Si-doped electron devices, which suggests surface states have a
similarly small impact on (100) C-doped hole devices. Nevertheless,
some non-linearity is seen in our (100) electron devices at low
$V_{g}$, which we tentatively assign to surface states (see, e.g.
figure~3 of \cite{BurkePRB12}). A study of sulfur passivation of
both electron and hole (100) devices with may be useful; one would
expect better passivation here since (100) only presents Ga
double-dangling bonds, which are robustly passivated by
sulfidization. The literature on passivation of (100) GaAs supports
this, but perfect passivation is not achievable, even for (100)
GaAs. One aspect to note is that a HfO$_{2}$ insulator between gate
and heterostructure surface is often used to prevent (apparent) gate
leakage in p-type AlGaAs/GaAs heterostructures~\cite{BurkePRB12,
CsontosAPL10}. This may adversely affect passivation treatment as
the organometallic precursors used in atomic layer deposition (ALD)
attack the GaAs surface~\cite{HinkleAPL08}. Although the relative
concentration of Ga-S bonds remained relatively unaffected in ALD
deposition of Al$_{2}$O$_{3}$ on (NH$_{4}$)$_{2}$S passivated (100)
GaAs~\cite{MilojevicAPL08}, it is unclear whether this would hold
for HfO$_{2}$ deposition. HfO$_{2}$ insulated C-doped (100)
AlGaAs/GaAs devices also display gate
hysteresis~\cite{CsontosAPL10}, but to a much lesser extent than
(311)A devices~\cite{BurkePRB12}. It would be interesting to study
the effect of sulfur passivation of C-doped (100) AlGaAs/GaAs
devices, as it may enable the highly-stable devices needed for
studying the fundamental physics of low-dimensional hole systems.

\section{Conclusions}
We studied the efficacy of (NH$_4$)$_2$S$_x$ surface passivation
treatment as a prospective solution to the gate hysteresis problem
in nanoscale devices made using (311)A AlGaAs/GaAs heterostructures.
XPS studies on (311)A and (100) heterostructures grown
simultaneously by MBE show very similar surface chemistry for both
surface orientations. PL measurements showed an improvement in PL
intensity by $40-65\times$ and $2-3\times$ for (100) and (311)A
surfaces, respectively, relative to untreated surfaces. The
comparative lack of PL intensity increase for (311)A is consistent
with a lack of reproducible improvement in gate hysteresis (311)A
FET devices made with the passivation treatment performed
immediately prior to gate deposition. We suggest that inefficacious
passivation, despite an obvious change in the surface chemistry,
arises due to the mixture of monovalent Ga and divalent As dangling
bonds present on the (311)A surface. We expect monovalent Ga to be
unpassivated by S -- giving at best a small increase in PL intensity
-- and divalent As-S bonds to be unstable, which could explain the
lack of reproducibility in gate characteristics for (311)A devices.
Further work on (311)A surface passivation is encouraged and could
include Cl- or Se-based treatments, or the addition of a GaS or
degenerately-doped GaAs cap layer as alternative passivation
strategies.

\ack This work was funded by Australian Research Council Grants
DP0877208, FT0990285 and DP110103802. DR and ADW acknowledge support
from DFG SPP1285 and BMBF QuaHL-Rep 01BQ1035. This work was
performed in part using the NSW node of the Australian National
Fabrication Facility (ANFF) and the Mark Wainwright Analytical
Center at UNSW. We thank Dr Bill Bin Gong for performing the XPS
measurements reported here, Tim Bray for assistance with some PL
measurements, and Tim Bray, Alex Hamilton and Oleh Klochan for
helpful discussions.

\appendix
\section{Practicality of incorporating (NH$_4$)$_2$S$_x$ passivation into device processing}
A range of possible sulfur treatment formulations exist, we focussed
on Na$_{2}$S and (NH$_{4}$)$_{2}$S based solutions as they provide
the best mix of ease and effectiveness. While Na$_{2}$S passivation
is reported to be more robust to light/oxygen than
(NH$_{4}$)$_{2}$S~\cite{SandroffJVSTB89}, treatment with
(NH$_{4}$)$_{2}$S gives complete oxide removal and a more sulfidized
surface with no traces of Na~\cite{SandroffAPL89, SpindtAPL89a}.
This behaviour can be enhanced by adding elemental S i.e., treating
with (NH$_{4}$)$_{2}$S$_{x}$~\cite{FanJJAP88, NannichiJJAP88}. Thus,
aqueous (NH$_4$)$_2$S and (NH$_4$)$_2$S$_{x}$ solutions are commonly
used for passivation of structures ranging from GaAs
metal-insulator-semiconductor field-effect transistors
(MISFETs)~\cite{JeongJJAP95} to InAs/GaSb photodiodes~\cite{LiAPL07}
to III-V nanowire devices~\cite{SuyatinNano07}.

To incorporate S passivation into device processing, it was
essential that treatment left the photoresist intact to facilitate
deposition of a photolithographically-defined gate such as in
figure~A1(a) and (c). Hence we began by studying the photoresist
compatibility of various sulfur passivation treatments including
aqueous Na$_{2}$S and (NH$_{4}$)$_{2}$S stock solutions diluted in
2-propanol, 2-methyl-2-propanol (t-butanol) and H$_{2}$O. We studied
three different resists: MicroChem S1813, a positive photoresist; AZ
nLOF2020, a negative photoresist; and MicroChem $950$k
polymethylmethacrylate (PMMA), a positive electron-beam lithography
(EBL) resist.

\begin{figure}
\includegraphics[width=8cm]{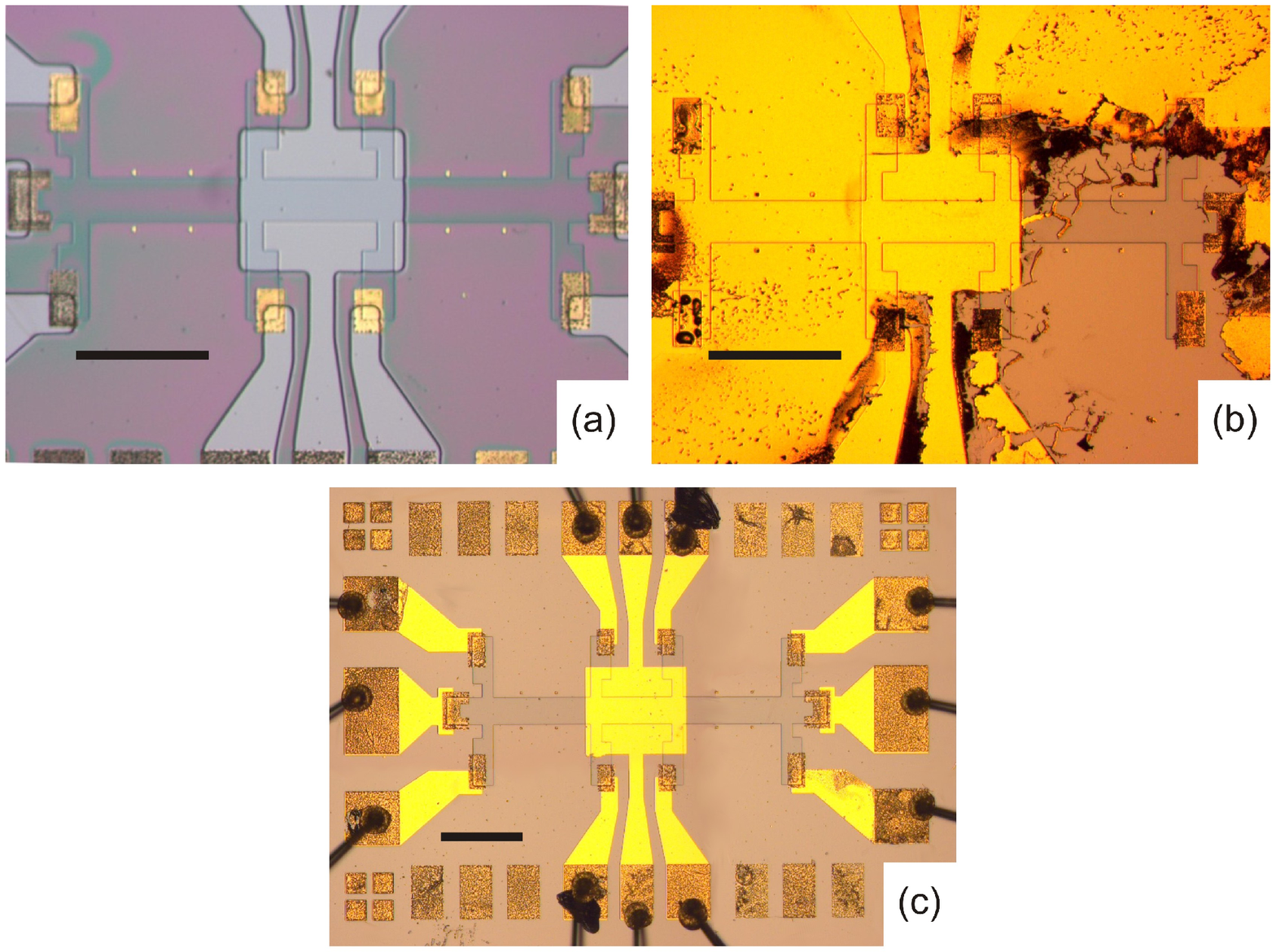}
\caption{Optical micrographs of a device (a) before passivation and
gate deposition and (b) after passivation with a 2-methyl-2-propanol
solution and gate deposition. (c) shows a successful device for
reference. The outline device center is a $130$~nm high hall bar
mesa, the grainy gold regions are annealed AuBe ohmic contacts and
the bright yellow regions are Ti/Au gates and leads. In (a), the
purple region is the developed resist. In (b), the gate metal is
poorly defined due to the photoresist being damaged by the alcohol
passivation solution. The black scale bars in all images represent
$300$~$\mu$m.}
\end{figure}

While alcoholic passivation is often more
effective~\cite{BesselovSemi98}, photoresist tends to be alcohol
soluble. A developed photoresist pattern began to dissolve within a
few seconds of immersion in a $2\%$ solution of (NH$_4$)$_2$S$_x$
stock solution in 2-propanol. The photoresist was completely removed
after $15$~s, a period insufficient for effective passivation of any
exposed GaAs. Similar results were obtained for stock solution
diluted in 2-methyl-2-propanol; an example where gate deposition was
attempted after passivation is shown in figure~A1(b). Clearly the
gate has not formed properly, with gate metal covering large regions
suffering unintentional photoresist removal during passivation. PMMA
proved more favorable due to its low solubility in 2-propanol.
Patterned PMMA films remained intact for immersions of up to
$15$~min in (NH$_4$)$_2$S$_x$ stock solution in 2-propanol. Although
this immersion time is sufficient for effective passivation,
electron-beam lithography is impractical for large area devices.
Qualitatively similar results to the above were obtained for
Na$_{2}$S solutions in 2-propanol and 2-methyl-2-propanol.

Turning now to purely aqueous solutions, we found a marked
difference between Na$_{2}$S and (NH$_4$)$_2$S$_x$ based solutions
for photoresist films. Photoresist films remained intact for at
least $5$~min. for (NH$_4$)$_2$S$_x$, whereas significant resist
damage resulted in $< 30$~s for Na$_{2}$S based solutions. Resist
damage was widespread for Na$_{2}$S solutions but concentrated at
the pattern edges. We suspect this occurs because Na$_{2}$S is a
much stronger base than (NH$_4$)$_2$S$_x$; UV-exposed photoresist is
normally developed in tetramethylammonium hydroxide (TMAH), which is
also basic. We performed the remainder of the studies in aqueous
(NH$_4$)$_2$S$_x$, since it was the only passivation solution for
which both photo- and EBL-resists remain intact for long immersions.

\section{Initial characterization of passivation solutions and their robustness}
After establishing the suitability of the aqueous (NH$_4$)$_2$S$_x$
solutions, we used PL measurements on (100) GaAs to confirm their
efficacy. The passivated samples used are listed in Table~B1.
Figure~B1(a) shows PL intensity versus wavelength for Sample B5
treated with the strong solution and Sample B6 treated with the weak
solution. The PL intensity is normalized to that obtained from an
otherwise equivalent untreated sample. A $\sim40\times$ and
$\sim90\times$ increase in PL intensity was observed for Samples B5
and B6, respectively. The $2.25\times$ improvement in PL intensity
produced by increasing passivation solution concentration by $\sim
200\times$ is small compared to the $40\times$ increase in PL
intensity produced using the weak solution. Regarding the appearance
of treated surfaces, Nannichi \etal~\cite{NannichiJJAP88} report
deposition of a thin off-white precipitate film on the GaAs surface
after (NH$_4$)$_2$S$_x$ passivation. We only obtain this when the
wafer is removed directly from the strong solution. This could be
prevented by diluting this solution with H$_{2}$O prior to removing
the sample. We found no appreciable change in PL intensity when
using this dilution method for preventing sulfide film formation.

\begin{table}
\centering \caption{List of samples discussed in the appendices. All are bulk, passivated
GaAs wafer pieces used for PL.}

\begin{tabular}{cccc}
\hline
Sample & Wafer & Surface & Passivation\\
\hline
B5 & bulk & (100) & strong\\
B6 & bulk & (100) & weak\\
B7 & bulk & (100)& strong\\
B8 & bulk & (100)& weak\\
B9 & bulk & (100)& weak + photoprocessed\\
\hline \hline
\end{tabular}
\end{table}

\begin{figure}
\includegraphics[width=8cm]{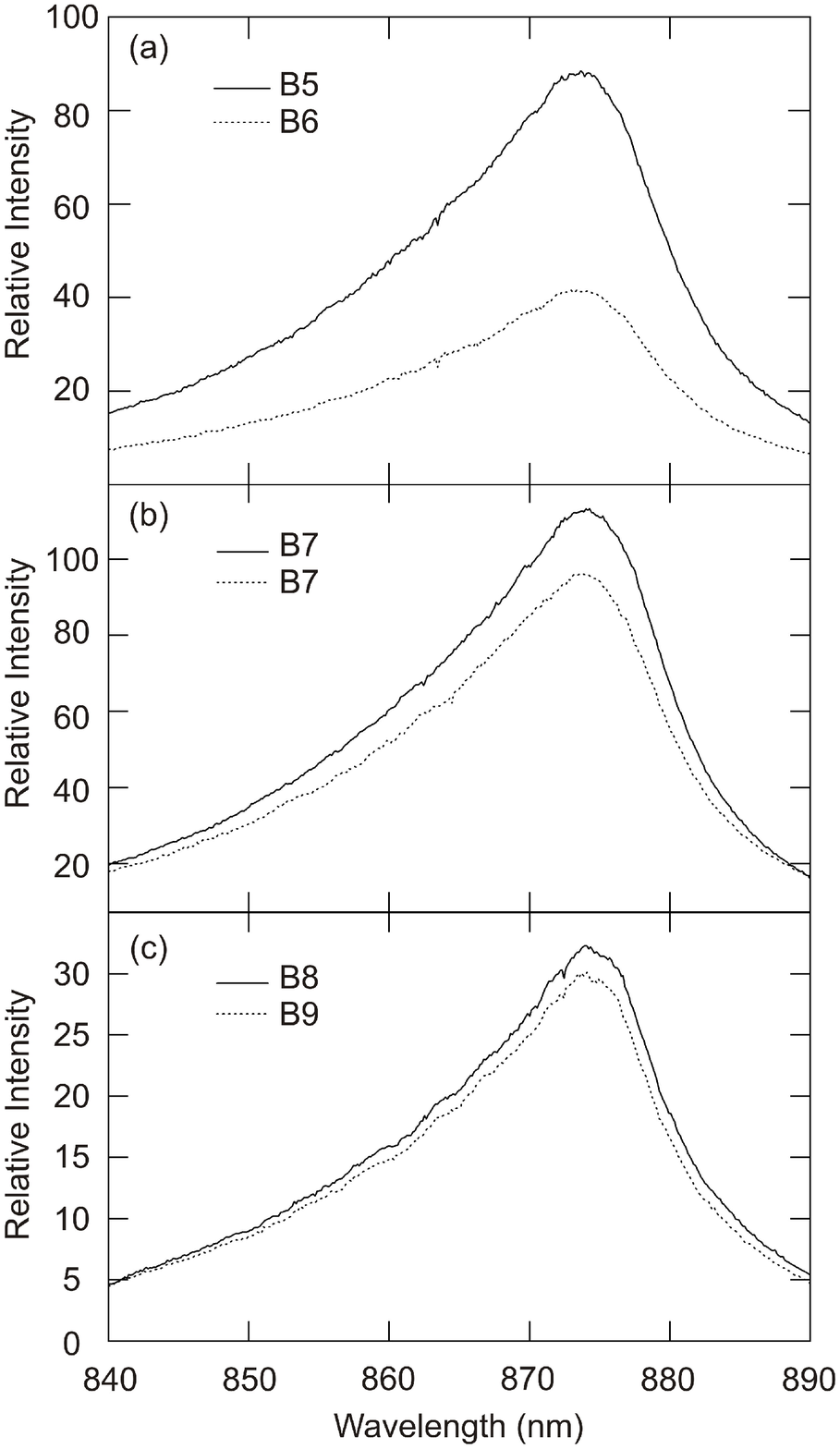}
\caption{(a) PL intensity versus wavelength for Samples B5 and B6.
Yields are normalized using the PL spectrum obtained from an
otherwise equivalent untreated sample. (b) PL spectrum of Sample B7
obtained immediately after passivation (solid line) and after one
week (dashed line) of storage in the dark under room atmosphere
conditions. (c) PL spectrum of Samples B8 and B9 showing that
comparable PL intensities are obtained whether or not the sample is
photoprocessed after passivation.}
\end{figure}

We also used PL to study the robustness of the treatment under
different storage conditions. Figure~B1(b) shows that the PL
intensity for B7 was only $15\%$ lower after 1 week stored in the
dark under room atmosphere conditions than immediately after
passivation. This is within the experimental error for a PL
measurement, suggesting little/no degradation of the passivation.
This confirms the brief air exposure between passivation and gate
deposition for devices D2-D5 is unlikely to adversely affect
performance. It can be desirable to thermally anneal the sample to
$\sim 350^{\circ}$C after passivation; patterned resists are
generally destroyed at such temperatures. This would require
passivation and annealing prior to resist deposition, thus we also
studied the robustness of a passivated surface to subsequent
photolithographic processing. Samples B8 and B9 were passivated
together, we then deposited an AZ nLOF2020 resist film on B9, baked
it at $110^{\circ}$C for $60$~s and developed in TMAH. Sample B9 was
not irradiated with UV prior to development as the surface
underneath the gates is normally protected by the photomask during
exposure. The PL intensity was not reduced by the addition of
photoprocessing steps, as shown in figure~B1(c). This confirms
device D6 should remained passivated during gate
photoprocessing/deposition.

\section{Raw PL data of unpassivated (100) and (311)A GaAs surfaces}
\begin{figure}
\includegraphics[width=8cm]{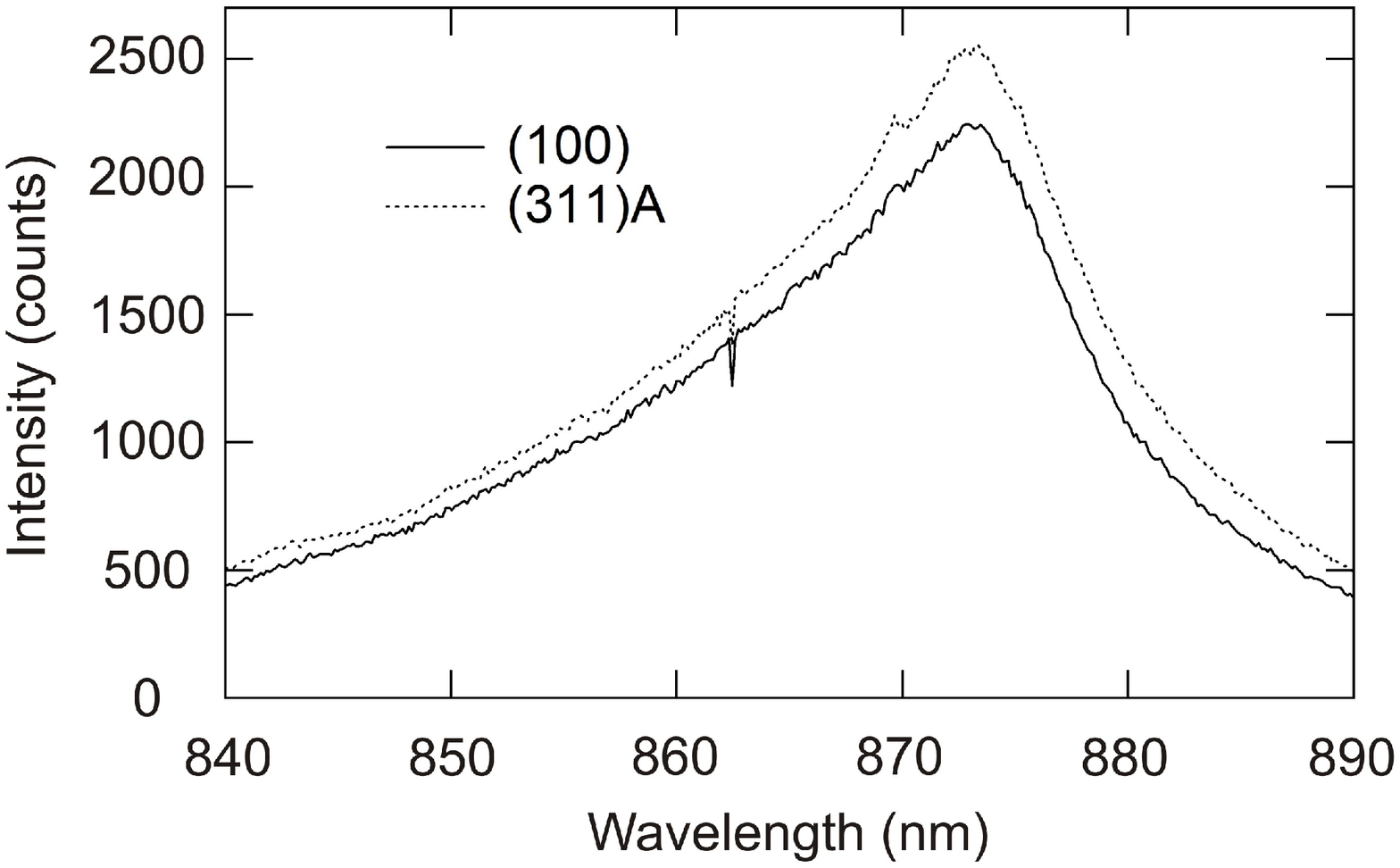}
\caption{PL intensity vs wavelength for the two reference samples in
section~3.2.}
\end{figure}

The data in figure~2(a/b) were normalized to the peak of untreated
(100)/(311)A oriented GaAs, respectively. Figure~C1 shows the raw
intensity for these reference samples. The data in figure~C1 has not
been normalized unlike all other presented PL data; the intensity
here refers to the recorded intensity incident on the CCD camera.
Note also that the (100) reference sample in figure~C1 is not the
same that in appendix~B; each experiment had its own unique
reference sample and this data is from reference samples used in
conjunction with the data in figure~2. Reference and passivated
samples were cleaved from adjacent positions on the host wafer to
ensure optimum similarity.

\end{document}